\documentclass[twocolumn]{aastex631}

\usepackage{subfigure}

\begin{document}

\title{Double-Shock Radiation in Supernova Wind Breakout and EP260321a}

\author{Zhuo Li}
\affiliation{Department of Astronomy, School of Physics, Peking University, Beijing 100871, China}
\affiliation{Kavli Institute for Astronomy and Astrophysics, Peking University, Beijing 100871, China}
\email{zhuo.li@pku.edu.cn}

\author{Rui-Ze Shi}
\affiliation{Department of Astronomy, School of Physics, Peking University, Beijing 100871, China}

\author{Qiang Wang}
\affiliation{Department of Astronomy, School of Physics, Peking University, Beijing 100871, China}

\begin{abstract}
We study the supernova (SN) shock breakout (SBO) from an optically thick stellar wind. The SN ejecta-wind interaction forms a forward shock (FS) plus reverse shock (RS) structure. The RS's internal energy and shock velocity are both a fraction $\mu\lesssim1/6$ of the FS's. The lower RS velocity makes it easier to build up local thermal equilibrium (LTE) and generate blackbody (BB) emission. In the same time a $\mu^{-1}\gtrsim6$ times larger non-BB emission may be produced by the likely non-LTE FS. We apply the theory to analyze the recent detected type Ic-BL SN2026gzf associated X-ray flash EP260321a, and find it is naturally accounted for by a SN wind breakout: the $\sim0.13$\,keV, thermal SBO emission can be produced by the RS during the interaction of a typical type Ic-BL SN ejecta ($E_k\sim10^{52}$erg and $M_e\sim1M_\odot$) and a stellar wind with a mass $M_w\sim10^{-4}M_\odot$ confined within a radius $r_w\sim10^{13}$cm; the X-ray rise time is mainly determined by the RS emission time, whereas the long duration and decrease time are due to the scatterings of SBO photons by a distant shell-like medium which has a large photosphere, $r_{ph}\sim10^{14}$\,cm; the early, within 30\,ks, optical emission can be naturally explained by the cooling emission of a shocked pure C/O envelope, since the observed SBO onset is delayed by $\gtrsim$ a few ks due to the large medium photosphere. BB+non-BB components from RS and FS, respectively, may be common in SN wind breakouts. SBO observations can be crucial in the study of progenitor stars by providing information for the ejecta from the SN explosion and the nearby wind from the pre-SN activities. 
\end{abstract}


\section{Introduction}
A supernova (SN) explosion generates a shock, which propagates in the progenitor star, heats, accelerates and ejects the stellar envelope. Given the large optical depth, the shock in the stellar interior is radiation-mediated and local thermal equilibrium (LTE) is established \citep{1976ApJS...32..233W}. When the radiation-mediated shock (RMS) propagates to a point that the optical depth of the stellar envelope in front of the shock drops to $\tau\sim c/v_s$, where $v_s$ is the shock velocity, the radiation can diffuse out and lead to a thermal UV/X-ray flash, so called supernova shock breakout (SBO) radiation \citep{1974ApJ...187..333C,1978ApJ...225L.133F,1978ApJ...223L.109K}. The thermal SBO radiation marks the first light of the SN explosion. After the SBO, a shock heated and expanding stellar envelope is left behind, and continue producing the so-called envelope cooling emission (CE) in UV/optical band in day scale \citep[e.g.,][]{1978ApJ...225L.133F,1992ApJ...393..742E,2011ApJ...728...63R,2007ApJ...667..351W,2008ApJ...683L.135C,2010ApJ...725..904N}. See \cite{2017hsn..book..967W} for a review of the SBO theory.

It is pointed out that the LTE may not be set up in a fast RMS where $v_s\gtrsim0.1c$, because the free-free emission cannot produce enough photons for a blackbody (BB) radiation \citep{2010ApJ...716..781K,1976ApJS...32..233W}. In such photon starving case, the radiation temperature is not a BB temperature and can be easily much higher, e.g., reaching an hard X-ray temperature for $v_s\gtrsim0.2c$  \citep{2010ApJ...716..781K}. Moreover, for trans-relativistic SBOs the radiation may even show a non-thermal spectrum \citep{2007ApJ...664.1026W,2010ApJ...716..781K,2012IAUS..279..274K}.

If the progenitor star is surrounded with dense and optically thick circum-stellar medium (CSM), the SN should break out from the CSM and generate the SBO radiation at the CSM breakout rather than stellar envelope breakout. This has been a new problem in the SBO study \citep{2006Natur.442.1008C,2007ApJ...664.1026W,2007MNRAS.375..240L,2007ApJ...667..351W,2008Natur.453..469S,2010ApJ...716..781K,2010ApJ...724.1396O,2011MNRAS.414.1715B,2012IAUS..279..274K,2019MNRAS.484.3502I,2025ApJ...993...46W}. 
There are increasing evidences that support the common existences of dense CSM or stellar winds around the progenitor stars of core collapse SNe, e.g., type II SNe \citep{2017NatPh..13..510Y,2018NatAs...2..808F} and type Ibc SNe \citep{2012ApJ...752...17W,2025MNRAS.542.2353C}.

In comparison, we notice some essential differences between envelope and CSM breakouts in the shock structure and dynamics. There commonly form a double-shock structure when two gases interact, including a forward shock (FS) and a reverse shock (RS), sweeping into two gasses respectively. In the envelope breakout, the FS accelerates with time when it is approaching the stellar surface and facing a sharp density gradient. Thus the FS departs further from the RS, if exists, with time. Usually only consider the SBO radiation generated by the FS. However, in the CSM/wind breakout, when the shock propagates from the stellar envelope into the dense CSM/wind, the interaction between the SN ejecta, formed by the shocked and ejected envelope material, and the CSM/wind forms a double-shock structure, with a FS sweeping up the CSM/wind material and a RS sweeping into the SN ejecta. As there is no steep density gradient in the CSM/wind, the FS is decelerating and the unshocked SN ejecta continue catching up; both FS and RS regions are compressed into thin shell. In the previous studies of CSM/wind breakouts, the SN ejecta is usually considered as a moving piston, and consider only the SBO radiation from the FS. Here we will investigate the double-shock structure and dyanmics, and the important role of the RS radiation on the SBO radiation in the CSM/wind breakouts.

SBO detections are rare because of their short duration and lacking of wide field-of-view (FOV) sky monitors in UV/X-ray bands. Previously only two X-ray flashes detected by Swift are reported to be likely SBO events, GRB060218 which is associted with a type Ic-BL SN2006aj \citep{2006Natur.442.1008C}, and XRF080109 associtated with a normal type Ib SN2008D \citep{2008Natur.453..469S}. In their spectra, GRB060218 shows a thermal component of $T\sim0.17$\,keV besides a dominant nonthermal component, whereas XRF080109 is better fitted with a nonthermal profile. They both show smooth light curves of a few hundreds to thousands seconds, consistent with SBO characteristics. Very recently, the wide FOV X-ray telescope, Einstein Probe (EP), reports the detection of a transient X-ray flash, EP260321a, that is associated with a type Ic-BL SN26gzf \citep{2026arXiv260610014Y}. EP260321a appears as a pure BB spectrum with temperature $T\sim0.13$\,keV in the 0.5-4keV band and a smooth X-ray light curve of a few ks. The thermal spectrum of EP260321a has been a strong feature that favors it as a SN Ic-BL associated SBO. We will apply our theoretical analysis to EP260321a in the context of CSM/wind breakouts.

In the following, Sec.\,\ref{sec:rms in wind} presents the theoretical investigation of the RMS structure and dynamics in the SN wind breakout, Sec.\,\ref{sec:EP260321a} is the analysis of the radiation processes of EP260321a in the context of wind breakouts, and finally Sec.\,\ref{sec:summary} gives the summary and discussion.

\section{RMS in ejecta-wind interaction}\label{sec:rms in wind}

\begin{figure}[t]
    \centering
    \includegraphics[width=0.9\linewidth]{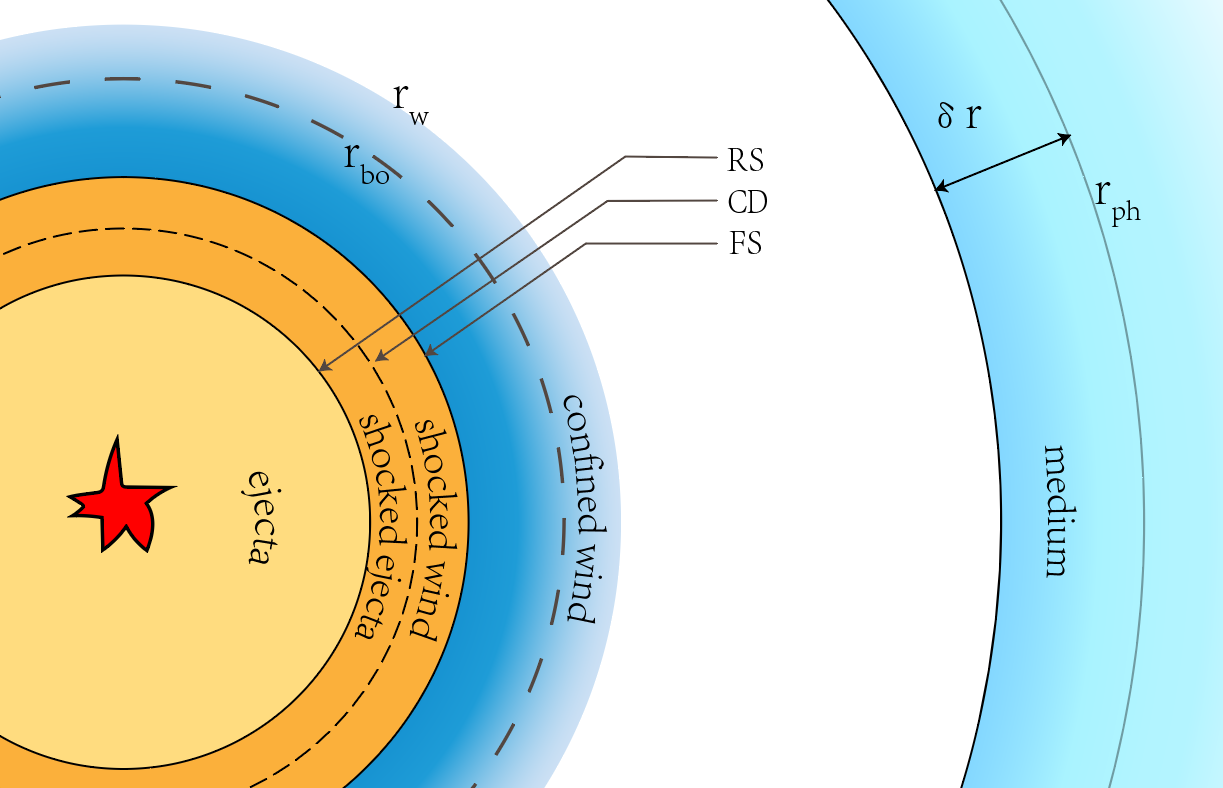}
    \caption{Schematic description of the geometry of the ejecta-wind interaction in a SN wind breakout and the outer extended medium, which is important for radiation transfer for EP260321a.}
    \label{fig:pic}
\end{figure}

In a successful SN explosion, the RMS should break out from the stellar surface. However, if the CSM is optically thick enough, the radiation in the radiation cannot escape faster than the shock; the shock should further break out from the CSM and generate the SBO radiation there. The dense CSM can be produced by pre-explosion strong stellar winds from the SN progenitor star. Thus the CSM can be a confined wind (see Sec.\,\ref{sec:EP260321a} that an extended wind does not work for EP260321a). Let us consider the ejecta-wind interaction and the generated SBO radiation. A schematic plot of the geometry of a SN wind breakout is shown in Figure~\ref{fig:pic}, also shown is an extended medium that is important in shaping the EP260321a's light curve (Sec.\,\ref{sec:EP260321a}).

\subsection{Double-shock structure}

The ejecta-wind interaction forms a double-shock structure, including a FS sweeping up the wind mass and a RS penetrating through the ejecta. Between the shocked wind mass and shocked ejecta mass there is a contact discontinuity. The shocked material can be compressed by the shocks into \emph{thin shells}. Let us denote regions 1--4 as the regions of the preshock wind, the shocked wind, the shocked ejecta and the preshock ejecta, respectively.

As the postshock regions are supposed to be dominated by radiation, we take the adiabatic index $\gamma=4/3$ for the postshock gas in both FS and RS regions. Consider that the preshock gas is cold for both RS and FS, i.e., the preshock pressure is $p_1=p_4=0$. From Rankine–Hugoniot condition, the postshock velocity, density and pressure are, respectively, given by
\begin{equation}\label{eq:FS jump}
    v_2=\frac67v_f,~~\rho_2=7\rho_1,~~p_2=\frac67\rho_1v_f^2
\end{equation}
for the FS and 
\begin{equation}\label{eq:RS jump}
    v_4-v_3=\frac67\bar{v}_r,~~\rho_3=7\rho_4,~~p_3=\frac67\rho_4\bar{v}_r^2
\end{equation}
for the RS. Here $Q_i$ with $i=$1--4 denotes quantities of different regions; for $i=1,4$ the quantities should be that right upstream of the FS and RS, respectively; and $v_f$ and $\bar{v}_r$ are the FS and RS velocities relative to the preshock wind and ejecta, respectively.
Take  
\begin{equation}\label{eq:FS-RS relation}
    p_2=p_3,~~v_2=v_3
\end{equation}
between FS and RS regions. Combining  Eqs.\,(\ref{eq:FS jump}), (\ref{eq:RS jump}), and (\ref{eq:FS-RS relation}), one derives  
\begin{equation}\label{eq:shock velocity}
    v_f=\frac76v_4(1+\mu )^{-1},~~\bar{v}_r=\mu v_f
\end{equation}
where we define $\mu\equiv(\rho_1/\rho_4)^{1/2}$.

Approximately, the mass ratio between RS and FS is equal to the ratio of mass fluxes crossing the shock fronts, $M_3/M_2\approx\rho_4\bar{v}_r/(\rho_1v_f)=(\rho_4/\rho_1)^{1/2}=1/\mu$. As the postshock internal energy per mass is $e_m=p/[(\gamma-1)\rho]$, the total postshock internal energy in the FS is 
$E_2=M_2(3p_2/\rho_2)=(18/49)M_2v_f^2$. Similarly, for the RS, $E_3=(18/49)M_3\bar{v}_r^2=(18/49)\mu M_2v_f^2$. With Eq.\,(\ref{eq:shock velocity}), the postshock internal energy in the FS and RS are, respectively,
\begin{equation}\label{eq:internal energy}
    E_2=\frac12M_2v_4^2(1+\mu )^{-2},~~E_3=\mu E_2.
\end{equation}
It might be noted that by the Rankine–Hugoniot condition, the postshock internal energy density at FS is $e_2=p_2/(\gamma-1)=\frac12\rho_2v_2^2$, which means after volume-integration in the thin postshock region, $E_2=\frac12M_2v_2^2$, consistent with Eq.\,(\ref{eq:internal energy}). 

\subsection{Shock dynamics}
In the free-expansion phase, consider that the kinetic energy distribution of the SN ejecta is $E_e(>v)=Kv^{-\delta}$, where $K=E_k v_e^{\delta}$,  $v_e=(2E_k/M_e)^{1/2}$, $E_k$ is the total kinetic energy, and $M_e$ is the total ejecta mass. The index is $\delta=(1+\frac1n-2\beta_1)/\beta_1$, where $\beta_1\simeq0.19$, and $n$ denotes the stellar density profile approaching the star radius $R_\star$ ($\rho\propto((R_\star-r)/R_\star)^n$) \citep{1999ApJ...510..379M,2010ApJ...716..781K}; usually, $n=3/2$ for convection dominated envelope, $n=3$ for radiation dominated envelope, and  $n=5$ in the critical case. If $n=3$, then $\delta\simeq5$. Assuming $n=3$, we take a fiducial value  $K=4.7\times10^{99}\mathcal{K}\,\rm (c.g.s)$, which corresponds to $E_k=10^{52}$ erg and $M_e=1M_\odot$. 

In the free expansion at late time, the radius of mass with velocity $v$ evolves with time as $r=vt$, then the separation for ejecta with velocity difference $dv$ is $dr=tdv=rdv/v$, and the spanned volume $dV=4\pi r^2dr=4\pi r^3dv/v$. The ejecta mass density is $\rho_e=dM_e/dV=(v/4\pi r^3)|dM_e/dv|$. By $E_e(>v)=\frac12M_ev^2$, one has $|dM_e/dv|=2(\delta+2)E_e(>v)v^{-3}$, then the ejecta density at a certain time $t$ is
\begin{equation}
    \rho_e(r)=\frac{(\delta+2)}{2\pi r^3}\frac{E_e(>v)}{v^2}
\end{equation}
with $v=r/t$.

Consider that the stellar wind is confined within radius $r_w$, and that the mass density is $\rho_w=Ar^{-2}$, where $A=\dot{M}/4\pi v_w$, with $\dot{M}$ and $v_w$ the mass loss rate and velocity of the stellar wind. We take a fiducial value $A=5\times10^{14}\mathcal{A}\,\rm g\,cm^{-1}$, corresponding to $\dot{M}/v_w=10^{-5}M_\odot\rm yr^{-1}/(km\,s^{-1})$.

When the shock propagates to radius $r$, the FS sweeps up a wind mass of 
\begin{equation}\label{eq:M2}
    M_2=\int^r4\pi r^2\rho_wdr=4\pi Ar
\end{equation}
whereas the ejecta mass across the RS is those with velocity $v>v_4=r/t$, i.e., $M_3=M_e(>v_4)=2E_e(>v_4)/v_4^2$. Using $\rho_1=\rho_w(r)$ and $\rho_4=\rho_e(r=v_4t)$, we derive  $\mu\equiv(\rho_1/\rho_4)^{1/2}=(4\pi Ar/[(\delta+2)M_e(>v_4)])^{1/2}=(M_2/[(\delta+2)M_3])^{1/2}=(\mu /(\delta+2))^{1/2}$, then
\begin{equation}\label{eq:theta}
    \mu =(\delta+2)^{-1}.
\end{equation}
For $n\simeq3\,(5)$, $\delta\simeq5\,(4)$, and $\mu\simeq1/7\, (1/6)$.

Before SBO from the dense wind, the radiation cannot escape and the shock is adiabatic. For energy conservation, we write $E_e(>v_4)=E_2+E_3+\frac12(M_2+M_3)v_2^2$. Using relations $E_3/E_2=M_2/M_3=\mu $ and $E_2=\frac12M_2v_2^2$, we have
\begin{equation}\label{eq:energy conservation}
    E_e(>v_4)=E_2(2+\mu+\mu^{-1})=\frac1{2}M_2v_4^2\mu^{-1}
\end{equation}
which, with helps of Eqs.\,(\ref{eq:M2}) and (\ref{eq:theta}), gives the shock dynamics, e.g., $v_4$ vs. $r$ relation,
\begin{equation}\label{eq:dynmics}
    v_4=\left[\frac K{2\pi(\delta+2)A}\right]^{\frac{1}{\delta+2}}r^{-\frac{1}{\delta+2}}.
\end{equation}
It might be noted that given the thin-shell geometry, the separation between FS and RS is negligible compared to their radii, thus $r$ can be the radius of both FS and RS.

\subsection{Breakout radiation}
The optical depth of wind beyond radius $r$ is 
\begin{equation}\label{eq:optdepth_wind}
    \tau_w=\int_r^{r_w}\rho_w\kappa dr=A\kappa(r^{-1}-r_w^{-1})\equiv\xi\frac{ A\kappa}r
\end{equation} 
where $\kappa\approx0.2\,\rm cm^2g^{-1}$ is the scatting opacity of the wind material without hydrogen, and for convenience, we define $\xi(r_w)\equiv1-(r/r_w)$ to replace $r_w$. If  $r\ll r_w$ then $\xi\sim1$; if $r_w-r\ll r$ then $\xi\approx(r_w-r)/r\ll1$. We adopt the notion  $\xi=10^{-1}\xi_{-1}$. 

When the shock propagates to a point that the radiation diffuses faster than the shock, the radiation breaks out. This occurs at $\tau_w\simeq c/v_f$ \citep{1976ApJS...32..233W,1999ApJ...510..379M}, which, with Eq.\,(\ref{eq:optdepth_wind}), gives
\begin{equation}\label{eq:sbo point}
    v_f=\frac{c}{\xi A\kappa}r.
\end{equation}

Considering $n=3$ (then $\delta\simeq5$ and $\mu\simeq1/7$), and using Eqs.\,(\ref{eq:dynmics}) and (\ref{eq:sbo point}), we derive the shock radius and velocities at SBO, 
\begin{equation}\label{eq:sbo radius}
    r_{bo}=4.1 \times 10^{12} \, \mathcal{A}^{0.75} \, \xi_{-1}^{0.88} \, \mathcal{K}^{0.12}\rm cm,
\end{equation}
\begin{equation}\label{eq:sbo velocity}
    v_{f,bo}=\mu ^{-1}\bar{v}_{r,bo}=1.2 \times 10^{10} \, \mathcal{A}^{-0.25} \, \xi_{-1}^{-0.12}\, \mathcal{K}^{0.12}\rm cm\,s^{-1}.
\end{equation}
Note, hereafter $\xi_{-1}=\xi(r_{bo};r_w)/10^{-1}$.
It is straightforward to obtain the internal energy and mass in the shocks at SBO by Eqs.\,(\ref{eq:internal energy}) and (\ref{eq:M2}),
\begin{equation}\label{eq:radiation energy}
    E_{3,bo}=\mu E_{2,bo}=2.0 \times 10^{47} \, \mathcal{A}^{1.25} \, \xi_{-1}^{0.63} \, \mathcal{K}^{0.37}\rm erg,
\end{equation}
\begin{equation}\label{eq:shock mass}
    M_{2,bo}=\mu M_{3,bo}=2.6 \times 10^{28} \, \mathcal{A}^{1.75} \, \xi_{-1}^{0.88} \, \mathcal{K}^{0.12}\rm g.
\end{equation}
We have also tried $n=5$ (then $\delta\simeq4$, and $\mu\simeq1/6$), and found the results only slightly change, e.g., the coefficients in Eqs.\,(\ref{eq:sbo radius}-\ref{eq:shock mass}) only change within 20-50\%.

If LTE is set up in both FS and RS (For non-LTE case, see discussion in the last paragraphs of this section), the radiation energy density is $e=aT^4=p/(\gamma-1)$, then, using Eqs.\,(\ref{eq:FS jump}) and (\ref{eq:sbo point}), the radiation temperature is derived, for both FS and RS,
\begin{equation}\label{eq:temperature}
    T_{bo}=1.1 \times 10^{6}\mathcal{A}^{-1/4} \, \xi_{-1}^{-1/2}\rm K,
\end{equation}
independent of the ejecta property parameters, $K$ and $n$.

We expect that a brief radiation flash with total energy $E_{bo}=E_{2,bo}+E_{3,bo}=E_{2,bo}(1+\mu)$ and temperature  $T_{bo}$ (if both FS and RS are in LTE) is generated at SBO.  
The SBO starts at the point when the FS arrives at $r_{bo}$, and lasts for a finite time, discussed here. Consider that the RMS is disrupted at $r_{bo}$ as radiation can escape from the shock. The shocked shells of regions 2 and 3 continue coasting with velocity $v_{2,bo}=v_2(r_{bo})$ without deceleration, keeping the same thicknesses, denoted $\Delta_2$ and $\Delta_3$\footnote{The radial spreading due to expansion in sound speed is negligible since the sound speed, comparable to thermal motion, is much smaller than $v_{2,bo}$.}. In the thin-shell approximation, $M_2=4\pi r_{bo}^2\Delta_2\rho_2$, we derive $\Delta_2\simeq r_{bo}/7$; similarly, $\Delta_3\simeq\mu r_{bo}/7$, thus $M_2\Delta_2\simeq M_3\Delta_3$. The emission duration should be determined by the photon diffusion. In small radii the photons are trapped, as the diffusion time is larger than the dynamical time $r/v_{2,bo}$. The photons can only escape around the point that the diffusion time is equal to the dynamical time, and the emission duration is estimated by the diffusion time then. For the FS emission, the emission duration is given on the one hand by $t_f\simeq \tau_2(r)\Delta_2/c$, with $\tau_2(r)=M_2\kappa/4\pi r^2$ being the diffusion time of FS photons, and on the other hand by $t_f\sim r/v_{2,bo}$. Canceling $r$ we have $t_f\sim[(A\kappa/7c)(r_{bo}/v_{2,bo})^2]^{1/3}$. Similarly, the RS emission duration is $t_r\simeq\tau_3\Delta_3/c+\tau_2\Delta_2/c\simeq2\tau_2\Delta_2/c$ ($\tau_3=M_3\kappa/4\pi r^2$), and we can derive $t_r\sim2^{1/3}t_f$. In total, the SBO duration is, with helps of Eqs.\,(\ref{eq:sbo radius}) and (\ref{eq:sbo velocity}),
\begin{equation}\label{eq:emission time}
    t_{bo}\simeq t_r\sim 530\, \mathcal{A} \, \xi_{-1}^{2/3} \rm s.
\end{equation}
The light crossing time of the breakout radius is $t_{cr}=r_{bo}/c<r_{bo}/v_{2,bo}<t_r$, so it does not determine the apparent emission duration. The SBO emission luminosity can be very roughly given by $L_{bo}\simeq E_{bo}/t_{bo}$. However, the photon propagation/diffusion in the outer/distant medium may still be important, and elongate the observed temporal profile (see Sec.\,\ref{sec:medium effect}).

One may expect that as the shocked region expands, the photons undergo adiabatic cooling before escape, say, $T\propto \rho^{1/3}\propto r^{-2/3}$. For the radius around which the photons diffuse out, $r\sim v_{2,bo}t_{bo}$, we have $r/r_{bo}\sim1.3\xi_{-1}^{-1/3}$, and the temperature may decrease to a fraction $\sim0.8\xi_{-1}^{2/9}$ of initial temperature, $T_{bo}$. Here we neglect this possible change of the temperature and energy of the apparent SBO emission.

We should check if the shocks keep being radiation-mediated at SBO. In the FS, a nucleus approaching the shock front from upstream may be decelerated by the radiation pressure $p_2$ on the nucleus-associated electrons. The change of the momentum per nucleon within a shock crossing time is $\Delta p\sim 0.5p_2\sigma_T\xi r_{bo}/v_{f,bo}$, where the factor 0.5 is roughly the electron number per nucleon in the material without H. At breakout, the momentum change per nucleon divided by its initial momentum $\sim m_pv_{f,bo}$ is
\begin{equation}\label{eq:rms condition}
    \eta_f=\frac{\Delta p}{m_p v_{f,bo}}\sim2.1 \mathcal{A}^{0.25}\xi_{-1}^{0.12}\mathcal{K}^{-0.12}.
\end{equation}
The insensitive dependence on parameters implies $\eta_f>1$ is valid in a wide parameter space, then the nucleus can be stopped by the radiation efficiently. In this case, the FS is still radiation mediated; it does not transits to a collision or collisionless shock before breakout. It is interesting that with Eq.\,(\ref{eq:sbo velocity}), \emph{the requirement of RMS, $\eta_f>1$, is equivalent to $v_{f,bo}\lesssim0.84c$}, independent of any other parameters. 
For the RS, in the same FS crossing time, as the pressure is the same and the shock velocity is smaller (Eq.\,\ref{eq:sbo velocity}), $m_p\bar{v}_r=m_pv_f\mu$, the ratio $\eta_r=\Delta p/m_p\bar{v}_r=\eta_f/\mu$ is much bigger, then the RS can be even more efficiently radiation-mediated than the FS.

It has been pointed out by \cite{2010ApJ...716..781K} that for RMS, if the shock velocity is fast, $v_s>0.1c$, the LTE cannot be established in the shock. This requirement is robust because it is very insensitive to preshock density. Thus, in comparison between RS and FS, the RS region is easier to satisfy LTE given lower shock velocity (Eq.\,\ref{eq:sbo velocity}), $\bar{v}_r=\mu v_f$, whereas the FS region easier for non-LTE. If the RMS is non-LTE, the apparent radiation temperature can be easily far larger than the LTE case, e.g., $T>10$\,keV for fast, $v_s>0.2c$, shocks \citep{2010ApJ...716..781K,2020MNRAS.492.1902I}, showing a thermal radiation in the hard X-ray range. For even faster shocks with $v_s>0.84c$, the shock may transit into collisionless shock before the SBO, thus the radiation could be non-thermal. It might be expected that in general a SN wind breakout produces a soft X-ray/UV BB component of smaller energy from the LTE RS, as well as a thermal, non-BB hard X-ray (or non-thermal) component of larger energy, by $\mu^{-1}$, from the non-LTE (or collisionless) FS. 

\section{EP260321a}\label{sec:EP260321a}
An X-ray transient, named EP260321a, was detected by EP/WXT, at a start time $T_0=$2026-03-21T12:16:08\,UTC \citep{2026arXiv260610014Y}. As the careful EP data analysis shows, the X-ray spectrum is well fit by a pure BB with a temperature $T_{th}\sim0.13$keV. Given the luminosity distance $D_L = 150\,\rm Mpc$, the time-integrated radiation energy in 0.3-10keV energy range is $\sim8.7\times10^{47}$\,erg, thus the total thermal energy is $E_{th}\sim1.5\times10^{48}$erg. The X-ray light curve in 0.4-2 keV lasts for a duration of $t_{dur}\sim2500$s, where a peak appears at a time $t_{rise}\sim700$\,s. A broad-lined type Ic SN, namely SN2026gzf, was detected associated with EP260321a, which first shows a bright optical emission within 10\,ks after $T_0$, whereas a SN optical emission at late times, $\gtrsim1$ day, similar to traditional SNe Ic-BL emission. 

With the thermal energy and temperature, if this is a BB temperature, the BB radius, i.e., the SBO radius, is
\[
    r_{bo}>\left(\frac{3E_{th}}{4\pi aT_{th}^4}\right)^{1/3}\sim2\times10^{12}\rm cm.
\]
However the progenitor stars of type Ic-BL SNe is believed to be massive Wolf-Rayet (WR) stars with the H and He envelope largely strippted  off and the star radius is $R_\star\lesssim1R_\odot$. The fact $r_{bo}\gg R_\star$ suggests that EP260321a should be a SBO event occurs in a dense CSM, rather than the stellar surface.

\subsection{Constraints from basic observables}
If the event is produced by SBO, the low observed temperature $T_{th}\ll 10$\, keV implies that the emission should be a BB radiation from a shock with low velocity $v_s<0.1c$, otherwise the temperature is very likely much higher, $T_{th}\gtrsim10$\,keV \citep{2010ApJ...716..781K}. So, in the wind-breakout case, the LTE should be established at least in the lower-speed RS region. 

If the thermal emission that EP observes is merely accounted for  by the RS radiation, then combining three equations, $T_{bo}=T_{th}$ (Eq.\,\ref{eq:temperature}), $E_{3,bo}=E_{th}$ (Eq.\,\ref{eq:radiation energy}) and $\bar{v}_{r,bo}=0.1c\,\bar{v}_{r,bo,0.1}$ (Eq.\,\ref{eq:sbo velocity}),  the parameters can be constrained by observed thermal energy and temperature and the LTE requirement, 
\begin{eqnarray}
    \mathcal{K}&=&77 E_{th,48} \bar{v}_{r,bo,0.1}^{5.0},\nonumber\\ 
    \mathcal{A}&=&1.5E_{th,48}^{2/3} T_{th,0.13}^{4/3}\bar{v}_{r,bo,0.1}^{-2},\\ 
    \xi_{-1}&=&0.44E_{th,48}^{-1/3} T_{th,0.13}^{-8/3}\bar{v}_{r,bo,0.1},\nonumber 
\end{eqnarray}
where $E_{th,48}=E_{th}/10^{48}\rm erg$, $T_{th,0.13}=T_{th}/0.13$\,keV, and  $\bar{v}_{r,bo,0.1}<1$ is required for LTE establishment.  On the other hand, if both FS and RS are in LTE, the EP observed thermal emission is contributed by both FS and RS, then the three equations becomes $E_{bo}\simeq8E_{3,bo}=E_{th}$, $T_{bo}=T_{th}$ and $v_{f,bo}=0.1c\,v_{f,bo,0.1}$, which derive 
\begin{eqnarray}
    \mathcal{K}&=&5.5\times10^{-4}E_{th,48} v_{f,bo,0.1}^{5.0},\nonumber\\
    \mathcal{A}&=&18E_{th,48}^{2/3} T_{th,0.13}^{4/3}v_{f,bo,0.1}^{-2},\\
    \xi_{-1}&=&0.13E_{th,48}^{-1/3} T_{th,0.13}^{-8/3}v_{f,bo,0.1},\nonumber
\end{eqnarray}
where $v_{f,bo,0.1}<1$. If both RS and FS are in LTE, then $\mathcal{K}\ll 1$. As $K(n=3)\propto E_k^{3.5}/M_e^{2.5}$, the small $K$ value suggests a low ejecta energy, $E_k\lesssim10^{51}$\,erg, and/or a large ejecta mass, $M_e\gtrsim10\,M_\odot$, and then a very low ejecta velocity $v_e\lesssim0.02c$. These may be somewhat abnormal for SN Ic-BL. The late time, $>1$\,day, supernova emission of SN2026gzf implies the ejecta energy and mass are comparable to our fiducial value $\mathcal{K}\sim1$. It is likely that $v_{f,bo}>0.1c$ hence the FS might not be in LTE situation to contribute to the EP observed, low temperature thermal emission.

In fact, it is likely that the rise time of the light curve is determined by the emission time, if the thermal X-rays are accounted for by the RS (see Sec.\,\ref{sec:medium effect} for detailed discussion). Thus we have one more equation from the observed rise time, $t_{bo}=t_{rise}$ (Eq.\,\ref{eq:emission time}), in addition to the two for thermal energy and temperature. All three model parameters can be solved, and the shock velocity is calculated for given observables:

\begin{eqnarray}
\mathcal{K} &=& 16  E_{th,48}^{2.7}T_{th,0.13}^{-1.7}  t_{rise,700}^{-3.8}, \nonumber\\
\mathcal{A} &=& 2.8  T_{th,0.13}^2  t_{rise,700}^{3/2}, \\
\xi_{-1} &=& 0.33T_{th,0.13}^{-3} t_{rise,700}^{-3/4}, \nonumber\\
v_{f,bo} &=& 1.5 \times 10^{10}  E_{th,48}^{0.33}T_{th,0.13}^{-0.33}  t_{rise,700}^{-0.75}\,\rm cm\,s^{-1}, \nonumber
\end{eqnarray}
where $t_{rise,700}=t_{rise}/700$\,s. The sensitive dependence on observables implies only orders of magnitude constraints of $\mathcal{K}$, $\mathcal{A}$ and $\xi_{-1}$. But the shock velocities are well constrained. The FS is non-LTE situation with $v_{f,bo}\sim0.5c$, whereas the RS has velocity $\bar{v}_{r,bo}\simeq v_{f,bo}/7\sim0.07c$ and may produce BB radiation accounting for the EP observed thermal emission.

In all cases, $\xi\ll1$ indicates that an extended stellar wind does not work for EP260321a. The wind should be confined with an edge at $r_w$, from which the shock breaks out, $r_{bo}\sim r_w$. The large $A$ value suggests a strong wind or abrupt ejection of mass before the SN explosion. The confinement radius $r_w\sim10^{13}$\,cm, together with an escape velocity of WR star $v_{esc}\sim10^3\rm \,km\,s^{-1}$, suggests a pre-SN ejection only $r_w/v_{esc}\sim$\, a few days before the SN explosion. Moreover, as $r_w\sim r_{bo}$, the ejected mass in the confined wind can be derived, $M_w\sim M_{2,bo}\sim10^{-4}M_\odot$.

\subsection{X-ray light curve: medium effect}
\label{sec:medium effect}

The observed X-ray duration is much longer than the emission time, $t_{dur}\gg t_{bo}$ (Eq.\,\ref{eq:emission time}). This implies that the observed light curve should be strongly affected by the medium where the photons propagate. Thus, the surrounding medium at $r>r_w$ should be still slightly optically thick, although the medium optical depth should be small, $\tau_m< c/v_f$, otherwise it will affect significantly the SBO properties, i.e., $E_{th}$ and $T_{th}$.

Suppose an extended medium beyond the confined wind results in a photosphere (or photon free-escaping sphere) at a radius $r_{ph}$. The optical depth between SBO radius and photospheric radius is then $\tau_m$. The inner-released photons undergoes random scattering by the medium before escape from the photosphere. Two timescales or effects are important for the observed temporal emission evolution by a distant observer. The first is the photon propagation time across the medium, $\sim\tau_m r_{ph}/c$ (This estimation is available for diffusion approximation when $\tau_m\gg1$). This may determine the rise time/peak time of the light curve. The second is the light crossing time of the photosphere extent, $t_{ph,cr}=r_{ph}/c$. For abrupt and isotropic emission from the photosphere, we can derive the observed flux or apparent luminosity evolution below. For a ring on the sphere, $(\theta,\theta+d\theta)$ with $\theta=0$ toward the observer, the observed emission energy is $dE_{obs}\propto\sin\theta d\theta\cos\theta$ and the time interval is $dt_{obs}=(r_{ph}/c)|d\cos\theta|=(r_{ph}/c)\sin\theta d\theta$. So the observed flux is $F\propto dE_{obs}/dt_{obs}\propto\cos\theta$. If define the observer time as $t_{obs}=(r_{ph}/c)(1-\cos\theta)$, i.e., $t_{obs}=0$ for emission at $\theta=0$, the observed flux drops linearly with time as 
\begin{equation}\label{eq: geometry eff}
    F(t_{obs})\propto 1 - \frac{t_{obs}}{t_{ph,cr}}.
\end{equation}
Thus, $t_{ph,cr}$ determines the decrease time of the light curve; the flux drops from the peak at $t_{obs}=0$ to zero at $t_{ph,cr}$. 

The above expectation for extended medium that the decrease time is smaller than the peak time by a factor of $\sim\tau_m$ is in contrast with observation, $t_{rise}<t_{dur}$. This implies that the outer medium may be shell-like, with a thickness  $\delta r<r_{ph}$, then the propagation time across the medium $\tau_m\delta r/c$ becomes smaller. In qualitative consideration, we should have the rise time determined by the larger one between the emission time and the shell propagation time, $t_{rise}\sim\max(t_{bo},\,\tau_m\delta r/c)$, and the decrease time or duration by the photosphere extent, $t_{ph,cr}\sim (t_{dur}-t_{rise})$. 

For $\tau_m \gtrsim1 $, the diffusion approximation is not valid. We therefore perform a simple Monte Carlo (MC) simulation to quantitatively calculate the X-ray light curve. We consider a shell-like medium with an outer radius $r_{ph}$, thickness $\delta r$ and scattering optical depth $\tau_m$. The photons are isotropically scattered with a mean free path of $\ell=\delta r/\tau_m$ in the shell. Given the spherical symmetry, the simulation becomes a 1D one: in time step $i$ a photon at radial position $r_i$ is randomly scattered into a new direction at polar angle $\theta_i$ with a free path $\ell_i$, then the step $i+1$'s position is calculated as $r_{i+1} = \sqrt{r_i^2 + \ell_i^2 + 2 r_i \ell_i\cos\theta_i}$. In the MC simulation, photons are injected at $r=r_{bo}$ along the radial direction, with a constant rate within a period of $(t_0,\,t_0+\Delta t_{inj})$, where $t_0=r_{bo}/v_{f,bo}$ is the SBO time since the SN explosion at $t=0$\footnote{The $t_0$ value does not affect the resulted light curve, because it is a constant included in both $t_j$ and $t_X$ and canceled out for $t_{obs,j}$ (Eq.\,\ref{eq:Xphoton-distribution}).}, and the injection time interval should be the SBO emission duration, $\Delta t_{inj}=t_{bo}$ (Eq.\,\ref{eq:emission time}). We calculate for each photon $j$ the time it crosses the photosphere, $t_j$, and its escaping polar angle, $\theta_{esc,j}$.
If a photon was scattered back into the inner region, we consider it absorbed or trapped by the SN ejecta. To account for the geometric effect of the photosphere, given the spherical symmetry, we calculate the arrival/observed time of photon $j$ for an observer,
\begin{equation}\label{eq:Xphoton-distribution}
    t_{obs,j} = t_j + \frac{r_{ph}}{c}(1 - \cos \theta_{esc,j})-t_X,
\end{equation}
where $t_X$ is the (source-frame) time since the SN explosion ($t=0$) when the first photons escape from the photosphere,
\begin{equation}\label{eq:t_X}
    t_X\simeq\frac{r_{bo}}{v_{f,bo}}+\frac{r_{ph}-r_{bo}}{c}= t_0+t_{ph,cr}-t_{cr}.
\end{equation} 
Repeating a large number of MC simulations, the distribution of $t_{obs,j}~(j=1,2,...N)$ gives us the observed light curves.

\begin{figure}[t]
    \centering
    \includegraphics[width=0.9\linewidth]{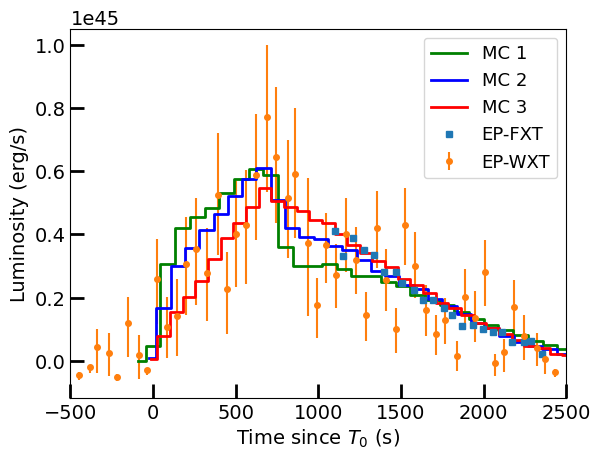}
    \caption{The prompt X-ray (0.4-2\,keV) light curve of EP260321a. The EP-WXT and EP-FXT data are taken from  \cite{2026arXiv260610014Y}. The curves are the model light curves from MC simulations: MC 1 (green), $\delta r = 4 \times 10^{12}\,\mathrm{cm}$, $\tau_m = 1.5$; MC 2 (blue), $\delta r = 2 \times 10^{12}\,\mathrm{cm}$, $\tau_m = 2$; MC 3 (red), $\delta r = 1.5 \times 10^{12}\,\mathrm{cm}$, $\tau_m = 3$. See text for the other parameters in the MC simulation. }
    \label{fig:X lc}
\end{figure}

Figure~\ref{fig:X lc} shows MC results in comparison with observation. The simulation setup is: the total energy of escaped photons is set as $E(\text{0.4-2\,keV})\simeq9\times10^{47} \,\mathrm{erg}$, the injection duration is $\Delta t_{inj}\simeq t_{rise}\approx700$\,s, the SBO radius is $r_{bo}\simeq5\times10^{12}$\,cm, and the photosphere radius is $r_{ph}\simeq c(t_{dur}-t_{rise})\simeq 6 \times 10^{13}$\,cm. The parameter sets for $\delta r$ and $\tau_m$ are presented in the caption of Figure~\ref{fig:X lc}. In general an initial peak shows up, which is contributed mainly by those photons directly penetrate through the medium shell without scatterings, then followed by a long decreasing tail, contributed by the medium-scattered photons and with the profile governed by the photospheric geometry extent (Eq.\,\ref{eq: geometry eff}).  
The results qualitatively catch the observed fast-rise-and-slow-decay evolution, with the rise time determined by the emission duration and the decrease time determined by the photosphere extent. The medium-shell mass can be estimated to be 
$M_m \sim4\pi r_{ph}^2  \tau_m / \kappa \sim 10^{-3}\tau_m\, M_{\odot}$.

\subsection{Early optical emission: envelope cooling}
\label{sec: optical}
The optical emission observed in late time, say $>1$\,day, can be largely explained by Ni-decay powered thermal emission of the SN ejecta, although additional contributions may be required \citep{2026arXiv260610014Y,2026arXiv260610002M,2026arXiv260609992O,2026arXiv260610011R}. However, the early optical emission observed $\sim30$\,ks within the X-ray start time \citep{2026arXiv260610014Y,2026arXiv260610009C} is not consistent with this interpretation. Different interpretations have been proposed \citep{2026arXiv260610014Y,2026arXiv260610009C}, but here we find that the CE from the shocked and expanding envelope can still be a natural explanation. 

First of all, we point out that the CE from the shocked wind and shocked ejecta is not important for the optical emission observed a few ks after $T_0$. At breakout, the total mass of the shocked shell, including FS plus RS (and the preshock mass up to $r_w$ is negligible as $\xi\ll1$), is $M_{23}=M_2+M_3\simeq8M_2$. As the shell expands its optical depth $\tau_{23}\sim M_{23}\kappa/4\pi r^2$ decreases quickly. It becomes optically thin at $r\gtrsim6\times10^{13}$\,cm, a distance comparable to $r_{ph}$ (Sec.\,\ref{sec:medium effect}). Thus when the shell expands beyond the photosphere, i.e., at observed times after the observed XBO start time $T_0$, the shocked shell becomes optically thin already and does not contribute and influence the CE any more.  
Here, we consider only the envelope CE from the SN ejecta not yet shocked by the RS.

As SN2026gzf is of type Ic-BL, with the H and He envelope largely stripped off from the progenitor star, we adopt the envelope cooling model for a C/O envelope from \citet{2017hsn..book..967W}; Eqs.\,(71) and (73) therein present the photospheric temperature $T_{ph}(t)$ and bolometric luminosity $L(t)$  as function of source-frame time $t$. The observed specific flux is given by
\begin{equation}
    F_\nu(t)=\frac{L(t)}{4\pi D_L^2} \frac{\pi B_{\nu}(T_{ph}(t))}{\sigma T_{ph}^4(t)},
\end{equation}
where $B_\nu(T_{ph})$ is the Planck function. The effect of the very small redshift of SN2026gzf is neglected here.

In the calculation of the envelope CE for optical data fitting, we fix the following model parameter values: the progenitor star's radius is $R_\star=1\,R_\odot$, typical for WR stars; the metal mass fraction $Z=1$ is taken for a pure C/O core without a He envelope; the stellar density profile related parameter, $f_\rho$, is simply fix to be $f_\rho=1$ as it is of order unity and has negligible effect on either $T_{\rm ph}(t)$ or $L(t)$. The remained important parameters are the SN ejecta's energy and mass, $E_k$ and $M_e$, which are relevant to $\mathcal{K}$.

Importantly, we should convert the observer time $t_{obs}$ since $T_0$ to the source-frame time $t$ by $t=t_{obs}+t_X$, where $t_X$,  as given in Eq.\,(\ref{eq:t_X}), depends on the other parameters. If including $t_X$ as a free parameter in addition to $E_k$, $M_e$, $\mathcal{A}$ and $\xi_{-1}$, we performed fitting to the early optical data, as well as the observables $E_{th}$ and $T_{th}$. It is found that a range of $10^3\,\text{s}<t_X<10^4$\,s is required for the fitting. \emph{The need of a significant $t_X$ value further supports the existence of a distant, dense CSM surrounding EP260321a/SN26gzf.} 

\begin{figure}[t]
    \centering
    \includegraphics[width=1\linewidth]{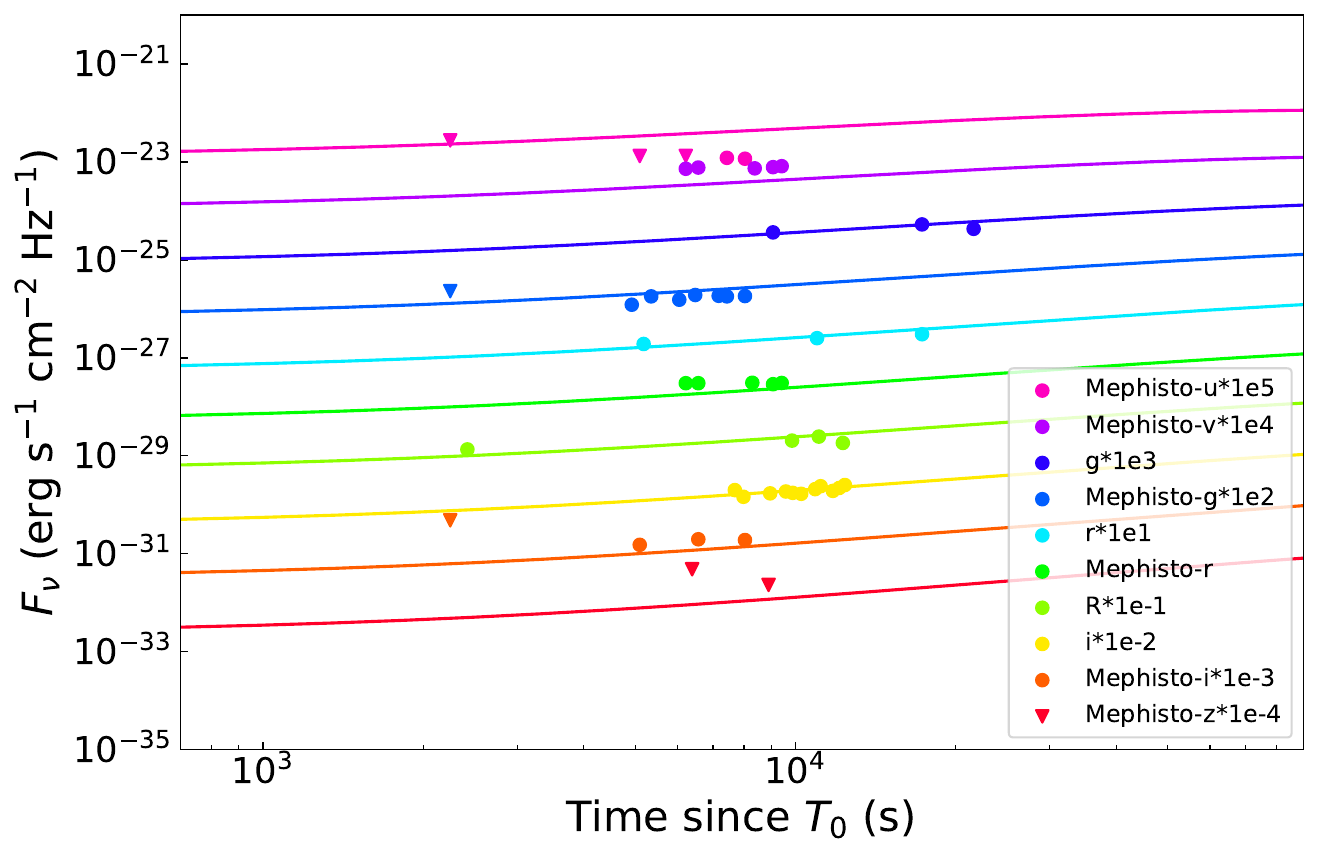}
    \caption{Multi-band early optical light curves of EP260321a/SN26gzf within $30$\,ks after the start time. The data are from \cite{2026arXiv260610014Y,2026arXiv260610009C}. The inverted triangles are the upper limits of flux. The curves are the multi-band light curves of the envelope cooling model. The adopted model parameter values are: 
    $E_k= 10^{52}\,\mathrm{erg}$, $M_e=1.1\,M_{\odot}$, $R_\star = 1R_\odot$, $Z = 1$, $f_\rho=1$, and $t_X=2400$\,s. }
    \label{fig:optical lc}
\end{figure}

\begin{figure}[t]
    \centering
    \includegraphics[width=1\linewidth]{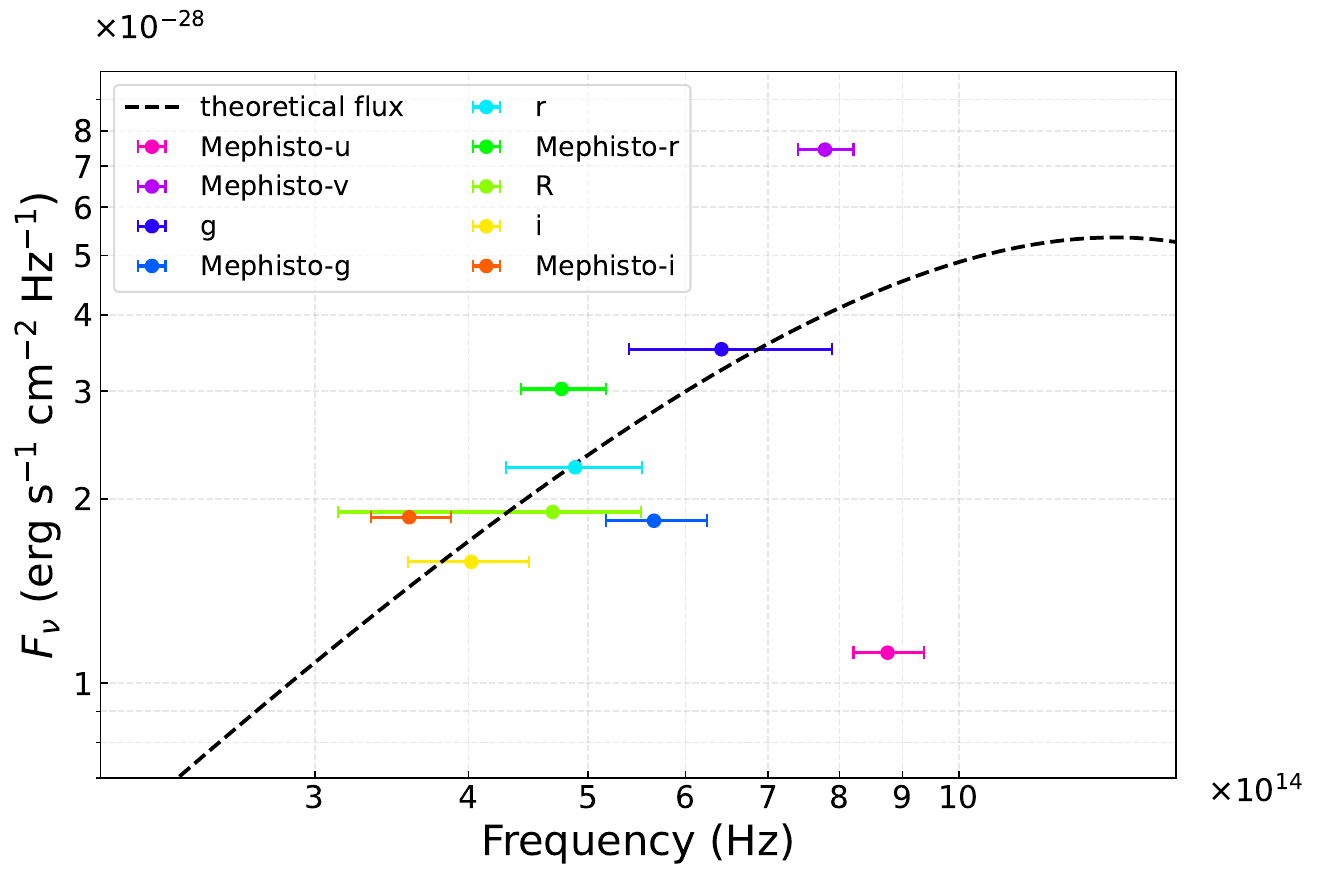}
    \caption{Optical spectrum of EP260321a/SN26gzf observed at $8500$\,s since the start time $T_0$. The model parameters are the same as in Figure~\ref{fig:optical lc}.}
    \label{fig:optical spectrum}
\end{figure}

Here, we show an example of modeling with $t_X\simeq2400$\,s, consistent with the picture setted up in Sec.\,\ref{sec:medium effect}, and  $E_k= 1.0\times 10^{52}\,\mathrm{erg}$ and $M_e=1.1\,\mathrm{M_{\odot}}$,  
comparable to those from late-time SN modeling \citep{2026arXiv260610002M} and \citet{2026arXiv260609992O}. 
Figure~\ref{fig:optical lc} shows the multi-band optical light curves at $t_{obs}\lesssim30$\,ks in comparison with the CE modeling. Figure~\ref{fig:optical spectrum} shows the spectrum of the envelope CE at $t_{obs}=8500\,\rm s$. We performed linear interpolation and extrapolation on the available multi-band data around this time, obtaining a rough spectrum. For the Mephisto data, the frequencies correspond to the central wavelengths of the bands \citep{2024ApJ...969..126Y}, while the frequency ranges are defined by the maximum and minimum wavelengths of each band. For the $gri$ bands\footnote{from https://svo2.cab.inta-csic.es/theory/fps/}, the frequencies refer to the effective wavelengths, with the ranges similarly given by the band wavelength limits.

It is seen that both the multi-band light curves (Figure~\ref{fig:optical lc}) and the snapshot spectrum (Figure~\ref{fig:optical spectrum}) can be well fitted by the envelope CE, unless some deviation in the Mephisto-$v$ and -$u$ bands. As shown in Figure~\ref{fig:optical spectrum}, these two bands exhibits first an excess followed by a sharp drop, compared with the model, probably due to some spectral line feature in the spectrum.

\section{Summary and discussion}\label{sec:summary}

We investigate the SN shock dynamics and the SBO properties when the SN shock breaks out from dense stellar wind. The main results are: the RS's velocity and radiation energy are smaller than those of the FS, respectively, by a factor of $\mu$ (Eq.\,\ref{eq:theta}); the shock velocities (Eq.\,\ref{eq:sbo velocity}), radiation energy (Eq.\,\ref{eq:radiation energy}), and temperature (Eq.\,\ref{eq:temperature}) of both FS and RS at breakout, as well as the SBO emission duration (Eq.\,\ref{eq:emission time}), as function of SN ejecta and wind properties are derived; the lower velocity of the RS makes it easier to satisfy the condition of RMS formation ($v_{f,bo}\lesssim0.84c$; Eq.\,\ref{eq:rms condition}) and the establishment of LTE and BB ($v_s\lesssim0.1c$). The BB plus non-BB components from RS and FS, respectively, may be general in SN wind breakout radiation.

We apply the theoretical results and analyze the nature of EP260321a under the context of wind breakout. The main conclusions on EP260321a/SN2026gzf are: the thermal X-ray radiation can be produced by the RS in ejecta-wind interaction, where the ejecta is with $E_k\sim10^{52}$erg and $M_e\sim1M_\odot$, whereas the wind is with a mass $ M_w\sim10^{-4}M_\odot$ confined within a radius $r_w\sim10^{13}$cm; the temporal X-ray profile should be shaped by both intrinsic and external factors, where the SBO emission duration determines the rise time and the large photosphere  of a slightly optically thick and extended ($r_{ph}\sim10^{14}$\,cm) medium shell determines the large decrease time; the early optical emission, observed about 10's ks within the X-ray start time, can be naturally explained by the CE of a shocked C/O envelope, provided that the observed SBO start time corresponds to a source-frame time later than the true SBO onset by a time of $t_X=t-t_{obs}\sim2$\,ks, which further requires an extended medium with a large photosphere (Eq.\,\ref{eq:t_X}).

Our theoretical results suggest that in general in the SN wind breakouts, the RS can establish LTE with $\bar{v}_{r,bo}<0.1c$ but the FS does not ($v_{f,bo}>0.1c$; Eq.\,\ref{eq:sbo velocity}). It is expected that a BB emission component in the SBO emission is generated by the RS, whereas a larger (by a factor of $\mu^{-1}$; Eq.\,\ref{eq:radiation energy}) energy of non-BB emission is produced by the FS. The non-BB emission from fast-speed RMS can still be thermal, but the temperature is much higher, e.g, $\gtrsim10$\,keV, determined by the electron-photon Compton scattering equilibrium \citep{2010ApJ...716..781K,2020MNRAS.492.1902I}. For even faster, trans-relativistic SBO, $v_f>0.84c$, the shock can not be mediated by radiation any more (see Eqs.\ref{eq:rms condition} and below), and may transits to collisionless shock before SBO, where non-thermal particles and radiation are expected to be produced. Moreover, in comparison the FS emission is somewhat shorter, with an emission time of $t_f\lesssim t_r$, whereas the RS emission is longer and determining the total SBO emission duration $t_{bo}\sim t_r$ (Eq.\,\ref{eq:emission time}), although the apparent emission duration may be modified by the medium scatterings (Sec.\,\ref{sec:medium effect}). In previous SBO events, the thermal plus non-thermal spectral components in GRB060218 \citep{2006Natur.442.1008C} may be explained by the RS plus FS SBO components. Moreover, the facts that the non-thermal emission shows much larger energy and shorter period \citep[e.g., Figs. 1 and 2 in][]{2006Natur.442.1008C} are all consistent with this double-shock emission picture. XRF080109's \citep{2008Natur.453..469S} non-thermal and low-energy ($\sim10^{46}$\,erg) properties can also be understood in this wind breakout picture, i.e., it may has lighter wind, $\mathcal{A}<1$, so that the FS is fast, $v_{f,bo}>0.8$, and the radiation energy is much smaller (Eqs.\,\ref{eq:sbo velocity} and \ref{eq:radiation energy}), and hence the RS's thermal emission could be too weak to observe.

In the analysis of EP260321a, we find that both the thermal X-ray and the early optical emission can be naturally and well explained by the SN wind breakout picture, where the required properties of the SN ejecta and the wind are typical for SNe Ic-BL, and consistent with the late-time SN2026gzf modeling. The important requirement for the optical modeling by the CE of shocked envelope is a shift of the onset time due to the delay by medium scatterings of the SBO photons. The required optically thick extended medium together with the dense confined wind suggests that the progenitor star is active in mass ejection/loss in timescales of day-month before the SN explosion, also in line of the observed variability of the progenitor star \citep{2026arXiv260610009C}. 
The soft X-ray band is dominated by a pure BB emission in EP260321a, but according to the prediction from our analysis, it is possible that there might be a large-energy emission from the FS in the hard X-ray band, $\gtrsim10$\,keV, far beyond the EP energy window. As the FS velocity may be a few$\times0.1c$, the FS emission component is expected to be thermal, with a temperature of 10's keV. A wider energy range, from soft to hard X-rays, is encouraged for future SBO observations of the RS+FS emission components. 

We did not consider here the reported temperature evolution of the X-ray emission in EP260321a, i.e., a slight change from $\sim0.13$\,keV to $\sim0.11$\,keV during the $\sim2$\,ks emission \citep{2026arXiv260610014Y}. However, one of the possible origins is the asymmetry in the CSM/wind, so that the radiation temperature may be angle dependent. If the photon temperature emitted at larger viewing angles from the SBO sphere is lower, these photons tend to escape from higher latitudes of the photosphere and arrive at observer later, by $\sim t_{ph,cr}\sim2$\,ks, leading to an apparent temperature decrease. Moreover, the effect of the temperature change on the observed flux is small. For the same photon number luminosity, the 0.4-2\,keV flux reduces between temperatures 0.13\,keV and 0.11\,keV by only a fraction of 32\%, smaller than the scatters and errors of the data (Figure~\ref{fig:X lc}).

\begin{acknowledgments}
This work is supported by National Key R\&D program of China (under the grant 2024YFA1611402), and Natural Science Foundation of China (No. 12575116).
\end{acknowledgments}

\bibliography{references}{}

@ARTICLE{2025ApJ...993...46W,
       author = {{Wasserman}, Tal and {Sapir}, Nir and {Szabo}, Peter and {Waxman}, Eli},
        title = "{The Optical to X-Ray Luminosity and Spectrum of Supernova Wind Breakouts}",
      journal = {\apj},
         year = 2025,
        month = nov,
       volume = {993},
       number = {1},
          eid = {46},
        pages = {46},
          doi = {10.3847/1538-4357/adfecc},
archivePrefix = {arXiv},
       eprint = {2504.00098},
 primaryClass = {astro-ph.HE},
       adsurl = {https://ui.adsabs.harvard.edu/abs/2025ApJ...993...46W}
}

@ARTICLE{2012ApJ...752...17W,
       author = {{Wellons}, Sarah and {Soderberg}, Alicia M. and {Chevalier}, Roger A.},
        title = "{Radio Observations Reveal Unusual Circumstellar Environments for Some Type Ibc Supernova Progenitors}",
      journal = {\apj},
         year = 2012,
        month = jun,
       volume = {752},
       number = {1},
          eid = {17},
        pages = {17},
          doi = {10.1088/0004-637X/752/1/17},
archivePrefix = {arXiv},
       eprint = {1201.5120},
 primaryClass = {astro-ph.HE},
       adsurl = {https://ui.adsabs.harvard.edu/abs/2012ApJ...752...17W}
}

@ARTICLE{2025MNRAS.542.2353C,
       author = {{Chiba}, Ryotaro and {Moriya}, Takashi J.},
        title = "{Hydrodynamic modelling of early peaks in Type Ibc supernovae with circumstellar interaction}",
      journal = {\mnras},
         year = 2025,
        month = sep,
       volume = {542},
       number = {3},
        pages = {2353-2362},
          doi = {10.1093/mnras/staf1382},
archivePrefix = {arXiv},
       eprint = {2504.06445},
 primaryClass = {astro-ph.HE},
       adsurl = {https://ui.adsabs.harvard.edu/abs/2025MNRAS.542.2353C}
}

@ARTICLE{1978ApJ...225L.133F,
       author = {{Falk}, S.~W.},
        title = "{Shock steepening and prompt thermal emission in supernovae.}",
      journal = {\apjl},
         year = 1978,
        month = nov,
       volume = {225},
        pages = {L133-L136},
          doi = {10.1086/182810},
       adsurl = {https://ui.adsabs.harvard.edu/abs/1978ApJ...225L.133F}
}

@ARTICLE{1974ApJ...187..333C,
       author = {{Colgate}, Stirling A.},
        title = "{Early Gamma Rays from Supernovae}",
      journal = {\apj},
         year = 1974,
        month = jan,
       volume = {187},
        pages = {333-336},
          doi = {10.1086/152632},
       adsurl = {https://ui.adsabs.harvard.edu/abs/1974ApJ...187..333C}
}

@ARTICLE{1978ApJ...223L.109K,
       author = {{Klein}, R.~I. and {Chevalier}, R.~A.},
        title = "{X-ray bursts from type II supernovae.}",
      journal = {\apjl},
         year = 1978,
        month = aug,
       volume = {223},
        pages = {L109-L112},
          doi = {10.1086/182740},
       adsurl = {https://ui.adsabs.harvard.edu/abs/1978ApJ...223L.109K}
}

@ARTICLE{2007MNRAS.375..240L,
       author = {{Li}, Li-Xin},
        title = "{Shock breakout in Type Ibc supernovae and application to GRB 060218/SN 2006aj}",
      journal = {\mnras},
         year = 2007,
        month = feb,
       volume = {375},
       number = {1},
        pages = {240-256},
          doi = {10.1111/j.1365-2966.2006.11286.x},
archivePrefix = {arXiv},
       eprint = {astro-ph/0605387},
 primaryClass = {astro-ph},
       adsurl = {https://ui.adsabs.harvard.edu/abs/2007MNRAS.375..240L}
}

@ARTICLE{1992ApJ...393..742E,
       author = {{Ensman}, Lisa and {Burrows}, Adam},
        title = "{Shock Breakout in SN 1987A}",
      journal = {\apj},
         year = 1992,
        month = jul,
       volume = {393},
        pages = {742},
          doi = {10.1086/171542},
       adsurl = {https://ui.adsabs.harvard.edu/abs/1992ApJ...393..742E}
}

@ARTICLE{2011ApJ...728...63R,
       author = {{Rabinak}, Itay and {Waxman}, Eli},
        title = "{The Early UV/Optical Emission from Core-collapse Supernovae}",
      journal = {\apj},
         year = 2011,
        month = feb,
       volume = {728},
       number = {1},
          eid = {63},
        pages = {63},
          doi = {10.1088/0004-637X/728/1/63},
archivePrefix = {arXiv},
       eprint = {1002.3414},
 primaryClass = {astro-ph.HE},
       adsurl = {https://ui.adsabs.harvard.edu/abs/2011ApJ...728...63R}
}

@ARTICLE{2010ApJ...725..904N,
       author = {{Nakar}, Ehud and {Sari}, Re'em},
        title = "{Early Supernovae Light Curves Following the Shock Breakout}",
      journal = {\apj},
         year = 2010,
        month = dec,
       volume = {725},
       number = {1},
        pages = {904-921},
          doi = {10.1088/0004-637X/725/1/904},
archivePrefix = {arXiv},
       eprint = {1004.2496},
 primaryClass = {astro-ph.HE},
       adsurl = {https://ui.adsabs.harvard.edu/abs/2010ApJ...725..904N}
}

@ARTICLE{2007ApJ...667..351W,
       author = {{Waxman}, E. and {M{\'e}sz{\'a}ros}, P. and {Campana}, S.},
        title = "{GRB 060218: A Relativistic Supernova Shock Breakout}",
      journal = {\apj},
         year = 2007,
        month = sep,
       volume = {667},
       number = {1},
        pages = {351-357},
          doi = {10.1086/520715},
archivePrefix = {arXiv},
       eprint = {astro-ph/0702450},
 primaryClass = {astro-ph},
       adsurl = {https://ui.adsabs.harvard.edu/abs/2007ApJ...667..351W}
}

@ARTICLE{2019MNRAS.484.3502I,
       author = {{Ioka}, Kunihito and {Levinson}, Amir and {Nakar}, Ehud},
        title = "{The spectrum of a fast shock breakout from a stellar wind}",
      journal = {\mnras},
         year = 2019,
        month = apr,
       volume = {484},
       number = {3},
        pages = {3502-3509},
          doi = {10.1093/mnras/stz270},
archivePrefix = {arXiv},
       eprint = {1810.11022},
 primaryClass = {astro-ph.HE},
       adsurl = {https://ui.adsabs.harvard.edu/abs/2019MNRAS.484.3502I}
}

@ARTICLE{2017NatPh..13..510Y,
       author = {{Yaron}, O. and {Perley}, D.~A. and {Gal-Yam}, A. and {Groh}, J.~H. and {Horesh}, A. and {Ofek}, E.~O. and {Kulkarni}, S.~R. and {Sollerman}, J. and {Fransson}, C. and {Rubin}, A. and {Szabo}, P. and {Sapir}, N. and {Taddia}, F. and {Cenko}, S.~B. and {Valenti}, S. and {Arcavi}, I. and {Howell}, D.~A. and {Kasliwal}, M.~M. and {Vreeswijk}, P.~M. and {Khazov}, D. and {Fox}, O.~D. and {Cao}, Y. and {Gnat}, O. and {Kelly}, P.~L. and {Nugent}, P.~E. and {Filippenko}, A.~V. and {Laher}, R.~R. and {Wozniak}, P.~R. and {Lee}, W.~H. and {Rebbapragada}, U.~D. and {Maguire}, K. and {Sullivan}, M. and {Soumagnac}, M.~T.},
        title = "{Confined dense circumstellar material surrounding a regular type II supernova}",
      journal = {Nature Physics},
         year = 2017,
        month = feb,
       volume = {13},
       number = {5},
        pages = {510-517},
          doi = {10.1038/nphys4025},
archivePrefix = {arXiv},
       eprint = {1701.02596},
 primaryClass = {astro-ph.HE},
       adsurl = {https://ui.adsabs.harvard.edu/abs/2017NatPh..13..510Y}
}

@ARTICLE{2007ApJ...664.1026W,
       author = {{Wang}, Xiang-Yu and {Li}, Zhuo and {Waxman}, Eli and {M{\'e}sz{\'a}ros}, Peter},
        title = "{Nonthermal Gamma-Ray/X-Ray Flashes from Shock Breakout in Gamma-Ray Burst-Associated Supernovae}",
      journal = {\apj},
         year = 2007,
        month = aug,
       volume = {664},
       number = {2},
        pages = {1026-1032},
          doi = {10.1086/519228},
archivePrefix = {arXiv},
       eprint = {astro-ph/0608033},
 primaryClass = {astro-ph},
       adsurl = {https://ui.adsabs.harvard.edu/abs/2007ApJ...664.1026W}
}

@ARTICLE{2008Natur.453..469S,
       author = {{Soderberg}, A.~M. and {Berger}, E. and {Page}, K.~L. and {Schady}, P. and {Parrent}, J. and {Pooley}, D. and {Wang}, X.-Y. and {Ofek}, E.~O. and {Cucchiara}, A. and {Rau}, A. and {Waxman}, E. and {Simon}, J.~D. and {Bock}, D.~C.-J. and {Milne}, P.~A. and {Page}, M.~J. and {Barentine}, J.~C. and {Barthelmy}, S.~D. and {Beardmore}, A.~P. and {Bietenholz}, M.~F. and {Brown}, P. and {Burrows}, A. and {Burrows}, D.~N. and {Byrngelson}, G. and {Cenko}, S.~B. and {Chandra}, P. and {Cummings}, J.~R. and {Fox}, D.~B. and {Gal-Yam}, A. and {Gehrels}, N. and {Immler}, S. and {Kasliwal}, M. and {Kong}, A.~K.~H. and {Krimm}, H.~A. and {Kulkarni}, S.~R. and {Maccarone}, T.~J. and {M{\'e}sz{\'a}ros}, P. and {Nakar}, E. and {O'Brien}, P.~T. and {Overzier}, R.~A. and {de Pasquale}, M. and {Racusin}, J. and {Rea}, N. and {York}, D.~G.},
        title = "{An extremely luminous X-ray outburst at the birth of a supernova}",
      journal = {\nat},
         year = 2008,
        month = may,
       volume = {453},
       number = {7194},
        pages = {469-474},
          doi = {10.1038/nature06997},
archivePrefix = {arXiv},
       eprint = {0802.1712},
 primaryClass = {astro-ph},
       adsurl = {https://ui.adsabs.harvard.edu/abs/2008Natur.453..469S}
}

@ARTICLE{2011MNRAS.414.1715B,
       author = {{Balberg}, Shmuel and {Loeb}, Abraham},
        title = "{Supernova shock breakout through a wind}",
      journal = {\mnras},
         year = 2011,
        month = jun,
       volume = {414},
       number = {2},
        pages = {1715-1720},
          doi = {10.1111/j.1365-2966.2011.18505.x},
archivePrefix = {arXiv},
       eprint = {1101.1489},
 primaryClass = {astro-ph.HE},
       adsurl = {https://ui.adsabs.harvard.edu/abs/2011MNRAS.414.1715B}
}

@ARTICLE{2006Natur.442.1008C,
       author = {{Campana}, S. and {Mangano}, V. and {Blustin}, A.~J. and {Brown}, P. and {Burrows}, D.~N. and {Chincarini}, G. and {Cummings}, J.~R. and {Cusumano}, G. and {Della Valle}, M. and {Malesani}, D. and {M{\'e}sz{\'a}ros}, P. and {Nousek}, J.~A. and {Page}, M. and {Sakamoto}, T. and {Waxman}, E. and {Zhang}, B. and {Dai}, Z.~G. and {Gehrels}, N. and {Immler}, S. and {Marshall}, F.~E. and {Mason}, K.~O. and {Moretti}, A. and {O'Brien}, P.~T. and {Osborne}, J.~P. and {Page}, K.~L. and {Romano}, P. and {Roming}, P.~W.~A. and {Tagliaferri}, G. and {Cominsky}, L.~R. and {Giommi}, P. and {Godet}, O. and {Kennea}, J.~A. and {Krimm}, H. and {Angelini}, L. and {Barthelmy}, S.~D. and {Boyd}, P.~T. and {Palmer}, D.~M. and {Wells}, A.~A. and {White}, N.~E.},
        title = "{The association of GRB 060218 with a supernova and the evolution of the shock wave}",
      journal = {\nat},
         year = 2006,
        month = aug,
       volume = {442},
       number = {7106},
        pages = {1008-1010},
          doi = {10.1038/nature04892},
archivePrefix = {arXiv},
       eprint = {astro-ph/0603279},
 primaryClass = {astro-ph},
       adsurl = {https://ui.adsabs.harvard.edu/abs/2006Natur.442.1008C}
}

@ARTICLE{2008ApJ...683L.135C,
       author = {{Chevalier}, Roger A. and {Fransson}, Claes},
        title = "{Shock Breakout Emission from a Type Ib/c Supernova: XRT 080109/SN 2008D}",
      journal = {\apjl},
         year = 2008,
        month = aug,
       volume = {683},
       number = {2},
        pages = {L135},
          doi = {10.1086/591522},
archivePrefix = {arXiv},
       eprint = {0806.0371},
 primaryClass = {astro-ph},
       adsurl = {https://ui.adsabs.harvard.edu/abs/2008ApJ...683L.135C}
}

@ARTICLE{2018NatAs...2..808F,
       author = {{F{\"o}rster}, F. and {Moriya}, T.~J. and {Maureira}, J.~C. and {Anderson}, J.~P. and {Blinnikov}, S. and {Bufano}, F. and {Cabrera-Vives}, G. and {Clocchiatti}, A. and {de Jaeger}, T. and {Est{\'e}vez}, P.~A. and {Galbany}, L. and {Gonz{\'a}lez-Gait{\'a}n}, S. and {Gr{\"a}fener}, G. and {Hamuy}, M. and {Hsiao}, E.~Y. and {Huentelemu}, P. and {Huijse}, P. and {Kuncarayakti}, H. and {Mart{\'\i}nez}, J. and {Medina}, G. and {Olivares E.}, F. and {Pignata}, G. and {Razza}, A. and {Reyes}, I. and {San Mart{\'\i}n}, J. and {Smith}, R.~C. and {Vera}, E. and {Vivas}, A.~K. and {de Ugarte Postigo}, A. and {Yoon}, S.-C. and {Ashall}, C. and {Fraser}, M. and {Gal-Yam}, A. and {Kankare}, E. and {Le Guillou}, L. and {Mazzali}, P.~A. and {Walton}, N.~A. and {Young}, D.~R.},
        title = "{The delay of shock breakout due to circumstellar material evident in most type II supernovae}",
      journal = {Nature Astronomy},
         year = 2018,
        month = sep,
       volume = {2},
        pages = {808},
          doi = {10.1038/s41550-018-0563-4},
archivePrefix = {arXiv},
       eprint = {1809.06379},
 primaryClass = {astro-ph.HE},
       adsurl = {https://ui.adsabs.harvard.edu/abs/2018NatAs...2..808F}
}

@INPROCEEDINGS{2012IAUS..279..274K,
       author = {{Katz}, Boaz and {Sapir}, Nir and {Waxman}, Eli},
        title = "{X-rays, {\ensuremath{\gamma}}-rays and neutrinos from collisionless shocks in supernova wind breakouts}",
    booktitle = {Death of Massive Stars: Supernovae and Gamma-Ray Bursts},
         year = 2012,
       editor = {{Roming}, P. and {Kawai}, N. and {Pian}, E.},
       series = {IAU Symposium},
       volume = {279},
        month = sep,
        pages = {274-281},
          doi = {10.1017/S174392131201304X},
archivePrefix = {arXiv},
       eprint = {1106.1898},
 primaryClass = {astro-ph.HE},
       adsurl = {https://ui.adsabs.harvard.edu/abs/2012IAUS..279..274K}
}

@ARTICLE{1976ApJS...32..233W,
       author = {{Weaver}, T.~A.},
        title = "{The structure of supernova shock waves.}",
      journal = {\apjs},
         year = 1976,
        month = oct,
       volume = {32},
        pages = {233-282},
          doi = {10.1086/190398},
       adsurl = {https://ui.adsabs.harvard.edu/abs/1976ApJS...32..233W}
}

@ARTICLE{2020MNRAS.492.1902I,
       author = {{Ito}, Hirotaka and {Levinson}, Amir and {Nagataki}, Shigehiro},
        title = "{Monte Carlo simulations of relativistic radiation-mediated shocks: II. photon-starved regime}",
      journal = {\mnras},
         year = 2020,
        month = feb,
       volume = {492},
       number = {2},
        pages = {1902-1913},
          doi = {10.1093/mnras/stz3591},
archivePrefix = {arXiv},
       eprint = {1910.08431},
 primaryClass = {astro-ph.HE},
       adsurl = {https://ui.adsabs.harvard.edu/abs/2020MNRAS.492.1902I}
}

@ARTICLE{2010ApJ...724.1396O,
       author = {{Ofek}, E.~O. and {Rabinak}, I. and {Neill}, J.~D. and {Arcavi}, I. and {Cenko}, S.~B. and {Waxman}, E. and {Kulkarni}, S.~R. and {Gal-Yam}, A. and {Nugent}, P.~E. and {Bildsten}, L. and {Bloom}, J.~S. and {Filippenko}, A.~V. and {Forster}, K. and {Howell}, D.~A. and {Jacobsen}, J. and {Kasliwal}, M.~M. and {Law}, N. and {Martin}, C. and {Poznanski}, D. and {Quimby}, R.~M. and {Shen}, K.~J. and {Sullivan}, M. and {Dekany}, R. and {Rahmer}, G. and {Hale}, D. and {Smith}, R. and {Zolkower}, J. and {Velur}, V. and {Walters}, R. and {Henning}, J. and {Bui}, K. and {McKenna}, D.},
        title = "{Supernova PTF 09UJ: A Possible Shock Breakout from a Dense Circumstellar Wind}",
      journal = {\apj},
         year = 2010,
        month = dec,
       volume = {724},
       number = {2},
        pages = {1396-1401},
          doi = {10.1088/0004-637X/724/2/1396},
archivePrefix = {arXiv},
       eprint = {1009.5378},
 primaryClass = {astro-ph.HE},
       adsurl = {https://ui.adsabs.harvard.edu/abs/2010ApJ...724.1396O}
}

@ARTICLE{1999ApJ...510..379M,
       author = {{Matzner}, Christopher D. and {McKee}, Christopher F.},
        title = "{The Expulsion of Stellar Envelopes in Core-Collapse Supernovae}",
      journal = {\apj},
         year = 1999,
        month = jan,
       volume = {510},
       number = {1},
        pages = {379-403},
          doi = {10.1086/306571},
archivePrefix = {arXiv},
       eprint = {astro-ph/9807046},
 primaryClass = {astro-ph},
       adsurl = {https://ui.adsabs.harvard.edu/abs/1999ApJ...510..379M}
}

@ARTICLE{2026arXiv260610011R,
       author = {{Rastinejad}, Jillian C. and {Srinivasaragavan}, Gokul and {Sarin}, Nikhil and {O'Dwyer}, Tanner and {Cenko}, S. Bradley and {Leung}, James K. and {Nugent}, Anya E. and {Perley}, Daniel A. and {Schroeder}, Genevieve and {Anand}, Shreya and {Ahumada}, Tomas and {Andreoni}, Igor and {Bochenek}, Aleksandra and {Corsi}, Alessandra and {Fremling}, Christoffer and {Ho}, Anna Y.~Q. and {Kasliwal}, Mansi M. and {Mo}, Geoffrey and {Salgundi}, Anirudh and {Sippy}, Kendall I. and {Sollerman}, J. and {Bellm}, Eric C. and {Chen}, Tracy X. and {Coughlin}, Michael W. and {Davis}, Michael C. and {De Colle}, Fabio and {Frostig}, Danielle and {Fryer}, Christopher L. and {Graham}, Michael J. and {Hall}, Xander J. and {Hinds}, K.-R. and {Izzo}, Luca and {Jacobson-Galan}, Wynn and {Lourie}, Nathan P. and {Maeda}, Keiichi and {Purdum}, Josiah and {Rusholme}, Ben and {Singh}, Avinash and {Stein}, Robert},
        title = "{A Multi-Wavelength View of the First Type Ic-BL Supernova with an Einstein Probe X-ray Shock Breakout}",
      journal = {arXiv e-prints},
         year = 2026,
        month = jun,
          eid = {arXiv:2606.10011},
        pages = {arXiv:2606.10011},
          doi = {10.48550/arXiv.2606.10011},
archivePrefix = {arXiv},
       eprint = {2606.10011},
 primaryClass = {astro-ph.HE},
       adsurl = {https://ui.adsabs.harvard.edu/abs/2026arXiv260610011R}
}

@ARTICLE{2010ApJ...716..781K
,
       author = {{Katz}, Boaz and {Budnik}, Ran and {Waxman}, Eli},
        title = "{Fast Radiation Mediated Shocks and Supernova Shock Breakouts}",
      journal = {\apj},
         year = 2010,
        month = jun,
       volume = {716},
       number = {1},
        pages = {781-791},
          doi = {10.1088/0004-637X/716/1/781},
archivePrefix = {arXiv},
       eprint = {0902.4708},
 primaryClass = {astro-ph.HE},
       adsurl = {https://ui.adsabs.harvard.edu/abs/2010ApJ...716..781K}
}

@ARTICLE{2026arXiv260610009C,
       author = {{Chen}, Ting-Wan and {Aryan}, Amar and {Yang}, Sheng and {Smartt}, Stephen J. and {Moriya}, Takashi J. and {Brennan}, Se{\'a}n J. and {Stritzinger}, Maximilian D. and {Martin}, Bailey and {Nicholl}, Matt and {Kong}, Albert K.~H. and {Gillanders}, James H. and {Dutta}, Anirban and {Schmidt}, Brian P. and {Cheng}, Yu-Chi and {Huber}, Mark E. and {Lai}, Cheng-Han and {Lee}, Chien-Hsiu and {Lee}, Yu-Hsing and {Ngeow}, Chow-Choong and {Smith}, Ken W. and {Ashall}, Christopher and {Auchettl}, Katie and {Burns}, Chris R. and {Chambers}, Kenneth C. and {Chen}, Zhi-Yue and {de Boer}, Thomas and {Hsiao}, Eric Y. and {Ngo Thanh Ho}, Khoa and {Hoogendam}, Willem B. and {Jones}, David O. and {Kankare}, Erkki and {Killestein}, Tom L. and {Kuncarayakti}, Hanindyo and {Lee}, Meng-Han and {Li}, Chuan-Jui and {Lin}, Chien-Cheng and {Lidman}, Christopher and {Lowe}, Thomas B. and {Magnier}, Eugene A. and {Medler}, Kyle and {M{\"o}ller}, Anais and {Moore}, Thomas and {Morrell}, Nidia and {Paek}, Gregory S.~H. and {Pfeffer}, Cameron M. and {Qiang}, Da-Chun and {Rauf}, Liana and {Reynolds}, Thomas M. and {Sankar. K}, Aiswarya and {Srivastav}, Shubham and {Tweddle}, Jack and {Wainscoat}, Richard and {Wang}, Ze-Ning and {Xiao}, Huangfei and {Zhu}, Zonghong},
        title = "{Decadal pre-explosion activity and circumstellar interaction in a supernova}",
      journal = {arXiv e-prints},
         year = 2026,
        month = jun,
          eid = {arXiv:2606.10009},
        pages = {arXiv:2606.10009},
          doi = {10.48550/arXiv.2606.10009},
archivePrefix = {arXiv},
       eprint = {2606.10009},
 primaryClass = {astro-ph.HE},
       adsurl = {https://ui.adsabs.harvard.edu/abs/2026arXiv260610009C}
}

@ARTICLE{2026arXiv260610014Y,
       author = {{Yuan}, Weimin and {Huang}, Qiu-Ju and {Zhu}, Jin-Ping and {Yu}, Yun-Wei and {Xu}, Dong and {Zhang}, Chen and {Li}, Zhuo and {Liu}, Yuan and {An}, Tao and {Gianfagna}, Giulia and {Zheng}, Weikang and {Du}, Guowang and {Liu}, Xing and {Jiang}, Ji-An and {Fynbo}, Johan P.~U. and {Pozanenko}, Alexei S. and {Jin}, Junjie and {Yang}, Yi and {Deng}, Jinsong and {Sun}, Hui and {Wu}, Guang-Lei and {Zhang}, Yu-Hao and {Wang}, Bao and {Wang}, Yu and {Wang}, Xiangyu and {Zhang}, Bin-Bin and {Chen}, Yong and {Zhang}, Yonghe and {Wang}, Bo and {Wang}, Xiaofeng and {Wu}, Xuefeng and {Dai}, Zigao and {An}, Jie and {Anupama}, G.~C. and {Balasubramanian}, Arvind and {Bao}, Congying and {Beri}, Aru and {Bhalerao}, Varun and {Brink}, Thomas G. and {Bruni}, Gabriele and {Cai}, Minxuan and {Cai}, Zhiming and {Chanchaiworawit}, Krittapas and {Chen}, Yehai and {Cheng}, Huaqing and {Cordier}, Bertrand and {Cui}, Chenzhou and {Cui}, Weiwei and {Dai}, Cuiyuan and {Eappachen}, D. and {Eselevich}, M.~V. and {Fan}, Xiao and {Fan}, Zhou and {Fang}, Yuan and {Feng}, Hua and {Filippenko}, Alexei V. and {Fu}, Shaoyu and {Gao}, He and {Geng}, Jinjun and {Goranskij}, Vitaly and {Guan}, Ju and {Han}, Dawei and {Hao}, Jinxin and {He}, Linbo and {He}, Min and {Hu}, Jingwei and {Huang}, Maohai and {Jia}, Shumei and {Jia}, Ziqing and {Jiang}, Shuaiqing and {Jin}, Chichuan and {Jin}, Ge and {Jonker}, Peter and {Klunko}, E.~V. and {Kong}, Albert K.~H. and {Li}, Chengkui and {Li}, Dongyue and {Li}, Rui-Zhi and {Li}, Wenxiong and {Liang}, Run-Duo and {Ling}, Zhixing and {Liu}, Congzhan and {Liu}, Huaqiu and {Liu}, Liangduan and {Liu}, Xiangkun and {Liu}, Xiaowei and {Liu}, Yuanqi and {Liu}, Zhengwei and {Lu}, Fangjun and {Mao}, Jirong and {Mao}, Xuan and {Moskvitin}, A.~S. and {Mu}, Haiyang and {Nandra}, Kirpal and {Ness}, Jan-Uwe and {Ni}, Kangrui and {Noysena}, Kanthanakorn and {O'Brien}, Paul and {Pan}, Haiwu and {Pan}, Yu and {Pankov}, N.~S. and {Piro}, Luigi and {Quirola-Vasquez}, J. and {Rau}, Arne and {Rea}, Nanda and {Sahu}, D.~K. and {Pawan Saikia}, Aditya and {Sanders}, Jeremy and {Song}, Liming and {Spiridonova}, Olga and {Sun}, Ning-Chen and {Sun}, Shengli and {Sun}, Xiaojin and {Tan}, Yuyin and {Thakur}, Aishwarya Linesh and {Tinyanont}, Samaporn and {Vlasyuk}, Valery and {Volnova}, A.~V. and {Wang}, Ailing and {Wu}, Hong and {Wu}, Qianrui and {Xu}, Haitao and {Xu}, Zelin and {Xue}, Changbin and {Yin}, Yi-Han Iris and {Zaznobin}, I.~A. and {Zhang}, Jia-Sen and {Zhang}, Shuang-Nan and {Zhang}, Songbo and {Zhang}, Yu and {Zhu}, Zipei and {Zou}, Zecheng and {Zhang}, Bing},
        title = "{Thermal X-rays breaking out from pre-explosion ejecta of a dying massive star}",
      journal = {arXiv e-prints},
         year = 2026,
        month = jun,
          eid = {arXiv:2606.10014},
        pages = {arXiv:2606.10014},
          doi = {10.48550/arXiv.2606.10014},
archivePrefix = {arXiv},
       eprint = {2606.10014},
 primaryClass = {astro-ph.HE},
       adsurl = {https://ui.adsabs.harvard.edu/abs/2026arXiv260610014Y}
}

@INCOLLECTION{2017hsn..book..967W,
       author = {{Waxman}, Eli and {Katz}, Boaz},
        title = "{Shock Breakout Theory}",
    booktitle = {Handbook of Supernovae},
         year = 2017,
       editor = {{Alsabti}, Athem W. and {Murdin}, Paul},
        pages = {967},
          doi = {10.1007/978-3-319-21846-5_33},
       adsurl = {https://ui.adsabs.harvard.edu/abs/2017hsn..book..967W}
}

@ARTICLE{2024ApJ...969..126Y,
       author = {{Yang}, Yuan-Pei and {Liu}, Xiangkun and {Pan}, Yu and {Er}, Xinzhong and {Liu}, Dezi and {Fang}, Yuan and {Du}, Guowang and {Cai}, Yongzhi and {Xu}, Xian and {Chen}, Xinlei and {Zou}, Xingzhu and {Guo}, Helong and {Liu}, Chenxu and {Cheng}, Yehao and {Kumar}, Brajesh and {Liu}, Xiaowei},
        title = "{Multiband Simultaneous Photometry of Type II SN 2023ixf with Mephisto and the Twin 50 cm Telescopes}",
      journal = {\apj},
         year = 2024,
        month = jul,
       volume = {969},
       number = {2},
          eid = {126},
        pages = {126},
          doi = {10.3847/1538-4357/ad4be3},
archivePrefix = {arXiv},
       eprint = {2405.08327},
 primaryClass = {astro-ph.HE},
       adsurl = {https://ui.adsabs.harvard.edu/abs/2024ApJ...969..126Y}
}

@ARTICLE{2026arXiv260609992O,
       author = {{O'Connor}, Brendan and {Hall}, Xander J. and {Busmann}, Malte and {Gruen}, Daniel and {Floris}, Alberto and {Cabrera}, Tomas and {Zhu}, Ziyuan and {Palmese}, Antonella and {Green}, Dylan and {Banovetz}, John and {Gassert}, Julius and {Fryer}, Christopher L. and {Ricci}, Roberto and {Troja}, Eleonora and {Shivaprasad}, Surya and {Zeimann}, Gregory R. and {Amsellem}, Ariel J. and {Bailey}, Stephen and {BenZvi}, Segev and {Dichiara}, Simone and {van Eerten}, Hendrik and {Hare}, Jeremy and {Hu}, Lei and {Irwin}, Christopher M. and {Kunnumkai}, Keerthi and {Malanchev}, Konstantin and {Maleki}, Mitra and {Moss}, Michael J. and {Myers}, Adam D. and {Pasham}, Dheeraj and {Ries}, Christoph and {Ryan}, Geoffrey and {Schlegel}, David and {Schmidt}, Michael and {Wilke}, Silona and {Yang}, Yu-Han},
        title = "{EP260321a/SN 2026gzf: The Faintest Shock Breakout Associated with a Broad-Lined Supernova}",
      journal = {arXiv e-prints},
         year = 2026,
        month = jun,
          eid = {arXiv:2606.09992},
        pages = {arXiv:2606.09992},
          doi = {10.48550/arXiv.2606.09992},
archivePrefix = {arXiv},
       eprint = {2606.09992},
 primaryClass = {astro-ph.HE},
       adsurl = {https://ui.adsabs.harvard.edu/abs/2026arXiv260609992O}
}

@ARTICLE{2026arXiv260610002M,
       author = {{Martin-Carrillo}, Antonio and {Th{\"o}ne}, Christina C. and {Leung}, James K. and {Corcoran}, Gregory and {de Ugarte Postigo}, Antonio and {Jonker}, Peter G. and {Izzo}, Luca and {Levan}, Andrew J. and {Gompertz}, Benjamin P. and {Basa}, St{\'e}phane and {Sarin}, Nikhil and {Quirola-V{\'a}squez}, Jonathan and {Eyles-Ferris}, Rob A.~J. and {Brivio}, Riccardo and {Watson}, Alan M. and {Cotter}, Laura and {Chac{\'o}n}, Jennifer Alexandra and {Rossi}, Andrea and {Melandri}, Andrea and {Kumnurdmanee}, Piramon and {Tanvir}, Nial R. and {Gupta}, Anshika and {Bauer}, Franz E. and {Ducoin}, Jean-Gr{\'e}goire and {Reguitti}, Andrea and {Misra}, Kuntal and {Xu}, Dong and {Vergani}, Susanna D. and {Fong}, Wen-fai and {Ackley}, Kendall and {Aguilar-Ruiz}, Edilberto and {Akl}, Dalya and {Aloy}, Miguel {\'A}ngel and {An}, Jie and {Angulo-Valdez}, Camila and {Antier}, Sarah and {Atteia}, Jean-Luc and {Becerra}, Rosa L. and {Breton}, Rene P. and {Butler}, Nathaniel R. and {Campana}, Sergio and {Carotenuto}, Francesco and {Casares Vel{\'a}zquez}, Jorge and {Chrimes}, Ashley A. and {D'Elia}, Valerio and {van Dalen}, Joyce N.~D. and {De Colle}, Fabio and {De Pasquale}, Massimiliano and {Dhillon}, Vik S. and {Dornic}, Damien and {Dyer}, Martin J. and {Ferro}, Matteo and {Fraser}, Morgan and {Fruchter}, Andrew S. and {Fortin}, Francis and {Galloway}, Duncan K. and {Garc{\'\i}a-Garc{\'\i}a}, Leonardo and {Geier}, Stefan and {Gill}, Ramandeep and {Globus}, No{\'e}mie and {Gualandi}, Roberto and {Guelfand}, Marion and {Guidolin}, Francesco and {Hartmann}, Dieter H. and {van Hoof}, Agnes P.~C. and {Jakobsson}, Pall and {Janghel}, Divyanshu and {Killestein}, Tom L. and {Klose}, Sylvio and {Kobayashi}, Shiho and {Kotak}, Rubina and {Kumar}, Amit and {Kuwata}, Asuka and {Laskar}, Tanmoy and {Lee}, William H. and {Lincetto}, Massimiliano and {Lombardi}, Gianluca and {L{\'o}pez-C{\'a}mara}, Diego and {Lyman}, Joseph D. and {Maiorano}, Elisabetta and {Maeda}, Keiichi and {Mandarakas}, Nikos and {Magnani}, Francesco and {Mao}, Jirong and {Moreno M{\'e}ndez}, Enrique and {Mar{\'\i}a Nicuesa Guelbenzu}, Ana and {Noysena}, Kanthanakorn and {Nuttall}, Laura K. and {O'Brien}, Paul T. and {O'Neill}, David and {Ochner}, Paolo and {Pereyra}, Margarita and {Pugliese}, Giovanna and {Ramsay}, Gavin and {Rhodes}, Lauren and {Saccardi}, Andrea and {Salvaterra}, Ruben and {S{\'a}nchez {\'A}lvarez}, Fredd and {Schneider}, Benjamin and {Schulze}, Steve and {Starling}, Rhaana L.~C. and {Steeghs}, Danny and {Ulaczyk}, Kzrysztof and {Ventura}, Chiara and {Zafar}, Tayyaba and {Zhu}, Zi-Pei},
        title = "{Failed jet breakout in the metal-poor broad-lined type Ic supernova 2026gzf}",
      journal = {arXiv e-prints},
         year = 2026,
        month = jun,
          eid = {arXiv:2606.10002},
        pages = {arXiv:2606.10002},
          doi = {10.48550/arXiv.2606.10002},
archivePrefix = {arXiv},
       eprint = {2606.10002},
 primaryClass = {astro-ph.HE},
       adsurl = {https://ui.adsabs.harvard.edu/abs/2026arXiv260610002M}
}
\bibliographystyle{aasjournal}
\end{document}